\documentclass[referee,useAMS,usenatbib,usegraphicx]{mn2e}
\usepackage{amssymb}

\newcommand{\cmt}[1]{}
\newcommand{\Rs}{R_{sub}}
\newcommand{\Rt}{R_{tot}}
\newcommand{\Ss}{S_{sub}}
\newcommand{\St}{S_{tot}}
\newcommand{\de}{\partial}
\newcommand{\pj}{p_{inj}}
\newcommand{\px}{p_{max}}
\newcommand{\degK}{\,\rmn{ K}}
\newcommand{\simgt}%
       {\,\hbox{\lower0.6ex\hbox{$\sim$}\llap{\raise0.6ex\hbox{$>$}}}\,}

   \newcommand{\simlt}%
       {\,\hbox{\lower0.6ex\hbox{$\sim$}\llap{\raise0.6ex\hbox{$<$}}}\,}

\title[Dynamical Feedback of Self-generated Magnetic Fields in Cosmic 
	Ray Modified Shocks]
{Dynamical Feedback of Self-generated Magnetic Fields in Cosmic Ray 	
	Modified Shocks}

\author[D. Caprioli, P. Blasi, E. Amato and M. Vietri]
{D. Caprioli$^{1}$\thanks{E-mail: d.caprioli@sns.it}, 
P. Blasi$^{2,3}$, E. Amato$^{3}$ and M. Vietri$^{1}$ \\
$^{1}$Scuola Normale Superiore, Piazza dei Cavalieri 1, 56100, Pisa, 
Italy\\
$^{2}$Fermilab, Center for Particle Astrophysics, Batavia, IL, USA\\
$^{3}$INAF-Osservatorio Astrofisico di Arcetri, Largo E. Fermi, 5, 
50125, Firenze, Italy\\}

\begin{document}
\date{\today}
\pagerange{\pageref{firstpage}--\pageref{lastpage}} \pubyear{2002}
\maketitle
\label{firstpage}

\begin{abstract}
We present a semi-analytical kinetic calculation of the process of non-linear
diffusive shock acceleration (NLDSA) which includes the magnetic field amplification
due to cosmic ray induced streaming instability, the dynamical reaction of the
amplified magnetic field and the possible effects of turbulent
heating. The approach is specialized to parallel shock waves and
 the parameters we chose are the ones appropriate to forward shocks
 in Supernova Remnants. Our calculation allows us to show that the
net effect of the amplified
magnetic field is to enhance the maximum momentum of accelerated particles
while reducing the concavity of the spectra, with respect to the standard
predictions of NLDSA. This is mainly due to the dynamical reaction of the
amplified field on the shock, which noticeably reduces the modification 
of the shock precursor. 
The total compression factors which are obtained for parameters typical of supernova
remnants are $R_{tot}\sim 7-10$, in good agreement with the values inferred from
observations. The strength of the magnetic field produced through excitation of
streaming instability is found in good agreement with the values inferred for
several remnants if the thickness of the X-ray rims are interpreted as due to
severe synchrotron losses of high energy electrons. We also discuss the
relative role of turbulent heating and magnetic dynamical reaction in
driving the reduction of the precursor modification. 
\end{abstract}


\maketitle

\section{Introduction}

The supernova remnant (SNR) paradigm for the origin of galactic cosmic
rays heavily relies on the mechanism for particle acceleration
being the \emph{Diffusive Shock Acceleration} (DSA) at the shock
front generated in the supernova blast. 
This mechanism, also known as \emph{first order Fermi acceleration}, has 
been studied in great detail and a two decades long work has led to 
the development of a kinetic non-linear theory that allows us to assess 
the importance of the dynamical reaction of the accelerated particles 
on the shock itself. The nonlinear effects described by the theory turn 
out to be not just corrections. They rather reveal the profound
reasons why the mechanism works in the first place as a cosmic ray
accelerator: the large efficiency required to explain the energetics 
($10-20\%$ of the kinetic energy of the blast wave going into
cosmic rays) and the large magnetic fields required to
explain the maximum energies observed in cosmic rays are the two main 
motivations for developing such a nonlinear theory, and, not
surprisingly, are among the most successful predictions of the theory. 

The initial attempt at building a nonlinear theory led to two-fluid
models \citep{dr_v80,dr_v81} which provided information on the
dynamics and thermodynamics of the shocked gas and cosmic ray gas
but not on the spectrum of the accelerated particles. A satisfactory 
kinetic approach, able to predict the spectrum of accelerated particles 
was later proposed by \cite{malkov1,malkov2} and
\cite{blasi1,blasi2}. More recently a kinetic model which takes into 
account the possibility of arbitrary diffusion coefficients was put forward by
\cite{amato1}. The self-excitation of unstable modes leading to magnetic field
amplification was then also introduced by \cite{amato2}. 

Parallel to these analytical approaches, developed primarily in the
assumption of quasi-stationarity of the acceleration process,
numerical approaches following the temporal evolution were also
developed \citep{bell87,je91,elli90,ebj95,ebj96,kj97,kj05,jones02}.
These have been of crucial importance for the description of some 
aspects of the phenomenology connected with the acceleration process 
in SNRs, especially during the Sedov-Taylor phase. Numerical
methods for the solution of the transport equation for cosmic rays 
and of the conservation equations for the plasma in the shock region 
have been extensively used to reproduce the observed multifrequency 
observations from single SNRs \citep[e.g.][]{V+05,cc+05}.

A typical result of all approaches to nonlinear shock acceleration is
that the spectra of accelerated particles are far from being power
laws. The concave shape leads to spectra as flat as $p^{-3.2}$ close
to the maximum momentum, corresponding to total compression factors
that may exceed $\sim 50-100$ \citep{amato1}. 
In general the total compression factor $R_{tot}$ is found to scale 
with the Mach number as $R_{tot}\propto M_0^{3/4}$ \citep{be99}. 
Such levels of shock modification do not compare well with some
observations, which suggest $R_{tot}\sim 7-10$ \citep[see e.g.][]{V+05}. 

From the phenomenological point of view this problem has been faced by invoking
some sort of turbulent heating: part (or all) of the energy in the form of
Alfv\'en waves which are responsible for the scattering of charged particles is
assumed to be damped on the thermal gas, thereby causing its heating in 
the upstream precursor region \citep{v-mck81,mck-v82}.
This process, originally investigated as a possible mechanism to limit
the magnetic field amplification, keeping the turbulent field in the
linear regime (e.g. $\delta B/B\ll 1$), is currently called upon also 
in situations in which $\delta B/B\gg 1$. 

\begin{table*}
\begin{minipage}{150mm}\label{tab:SNR}
\caption{Parameters for 5 well studied SNRs.}
\begin{tabular}{ccccc}
\hline
SNR & $u_0 (\rmn{km/s})$ & $B_2(\umu\rmn{G})$ & $\alpha_2\times 10^3$\\
\hline
Cas A& 5200 (2500)& 250--390 & 32 (36) \\
Kepler & 5400 (4500)& 210--340 &  23 (25)\\
Tycho & 4600 (3100)& 300--530 &  27 (31)\\
SN 1006 & 2900 (3200)& 91--110 & 40 (42) \\
RCW 86 & (800) & 75--145 & 14-35 (16-42)\\
\hline
\end{tabular}

\medskip
The value are from \cite{V+05} (in parentheses) and from \cite{P+06}.
In order to estimate the normalized downstream magnetic pressure 
$\alpha_{2}=B_2^2/(8\pi\rho_0u_0^2)$, for the
SNRs discussed by \cite{P+06} we used $\rho_0=0.1\,m_p/\rmn{cm}^3$ (SN 1006)
and $\rho_0=0.5\,m_p/\rmn{cm}^3$ (in the other cases). 
\cite{V+05}, instead, provide directly $\alpha_2$.
\end{minipage}
\end{table*}

An important piece of information has been recently added to the debate on
whether SNRs can be the sources of galactic cosmic rays: Chandra X-ray
observations of some remnants showed narrow filaments of non-thermal origin
\cite[see][and references therein for a review]{vink}.
If the thickness of the rims is assumed to be due to severe synchrotron
losses limiting the lifetime of high energy electrons, then one can infer 
the strength of the magnetic field in the downstream region, which turns 
out to be $\sim 100$ times stronger than magnetic fields in the ISM. 

On the other hand, it has been proposed that the narrow rims may reflect the
damping scale of the magnetic turbulence rather than the loss length of
electrons \citep{pohl05}. This interpretation is at odds with the 
morphology of the radio emission, as discussed by \cite{radio}  \citep[see also the
discussion in][]{morlino}, but at present it is not possible to rule
it out. If it turns out to be correct, then there would be no
observational constraint on the magnetic field in the shock region. 

If on the other hand the interpretation based on strong synchrotron losses
is confirmed, the inferred levels of magnetization can be interpreted as a
result of streaming instability induced by cosmic rays efficiently
accelerated at a parallel shock front
\citep{bell78,bl2001,bell2004}, although 
alternative mechanisms have also been put forward \citep[e.g.][]{gj07}. In
Table~\ref{tab:SNR} we list the SNRs where evidence has been collected, from
X-ray observations, for a strong magnetic field: $u_0$ is the shock velocity,
$B_2$ is the value of the magnetic field downstream of the shock as inferred
from the X-ray brightness profile and $\alpha_2$ is the magnetic energy density
immediately downstream of the forward shock, in units of the total
kinetic pressure at upstream infinity ($\alpha_{2}=B_2^2/(8\pi\rho_0u_0^2$)). 
The data are from \cite{V+05} (numbers in parentheses) and from \cite{P+06}.

The efficient acceleration and the magnetic field amplification are the two most
impressive manifestations of nonlinear diffusive acceleration at SNR shocks.
Both these aspects have been included in a Monte Carlo scheme by \cite{elli06},
with the magnetic field amplification described accordingly to the
phenomenological scenario proposed by \cite{bl2001}. \cite{elli06} find that,
when amplification is efficient, the wave pressure makes the plasma upstream of
the shock less compressible and the change in the energy density of the magnetic
turbulence across the subshock strongly affects $\Rs$, which in turn affects the
injection efficiency and the entire process of cosmic ray acceleration.

In a previous paper \citep{jumpl} we used a three-fluid approach (gas, 
cosmic rays and Alfv\'en waves), to show  the very general nature 
of the magnetic dynamical feedback: the shock dynamics is
significantly affected by the turbulence backreaction as soon as the 
magnetic pressure becomes comparable to that of the gas upstream of
the subshock. In particular, for the magnetization levels inferred
from available observations (see Tab.~\ref{tab:SNR}) and the
saturation values that are expected from different amplification
mechanisms proposed in the literature, we found that $\Rt$ would naturally 
lead to values in the range 6-10, in good agreement with the values currently 
inferred from observations. 

In the present paper we use a kinetic model for particle acceleration at a parallel shock in the non-linear regime, together with the
modified conservation equations in the
precursor and at the subshock in order to describe particle acceleration, the
dynamical reaction of accelerated particles, the generation of magnetic field
through streaming instability and the dynamical reaction of the magnetic field
on the shock, which also results in modified cosmic ray spectra.

The paper is structured as follows: in Sec.~\ref{sec:jumpcond}  we 
write down the correct jump conditions for a parallel, modified shock when 
magnetic turbulence is excited by the cosmic ray streaming, as in 
\cite{jumpl}.
In Sec.~\ref{sec:model} we summarize our kinetic model based on 
that by \cite{amato2}, including the self-consistent treatment of the 
magnetic field amplification via resonant streaming instability. The latter is described
in detail in Sec.~\ref{sec:resSI}. 
In Sec.~\ref{sec:results} we present our main results, namely the
solutions for DSA with resonant streaming instability: in particular we discuss the
reduced modification of the precursor and the consequent steepening of the 
spectrum near the (increased) maximum momentum. 
In Sec.~\ref{sec:TH} we investigate the combined effect of magnetic reaction
and turbulent heating and we show that the dominant effect on the precursor
modification is likely to be that of the magnetic reaction.
Our conclusions are in Sec.~\ref{sec:conclusions}.

\section{Dynamics of a magnetized cosmic ray modified shock}
\label{sec:jumpcond}

The pressure of the accelerated particles upstream of the shock surface
leads to the formation of a shock precursor, in which the fluid speed
gradually decreases while approaching the shock. One can describe this
effect by introducing two compression factors $R_{tot}=u_0/u_2$ and
$R_{sub}=u_1/u_2$, where $u$ is the fluid velocity and the indexes 
'$0$', '$1$' and '$2$' refer, here and in the following, to quantities 
taken at upstream infinity ($x=-\infty$), and immediately
upstream ($x=0^-$) and downstream ($x=0^+$) of the subshock, respectively.

We consider a non-relativistic plane shock whose normal is parallel to the background magnetic field $B_0$.
The equations defining the jump conditions at the shock surface in the
presence of cosmic rays and self-generated Alfv\'en waves can be
written as
\begin{equation}
	\left[\rho u \right]_1^2=0\;,
\end{equation}
\begin{equation}
	\left[\rho u ^2+p+p_{w}\right]_1^2=0\;,
\end{equation}
\begin{equation}
	\left[\frac 1 2\rho u^3+\frac{\gamma}{\gamma-1}up
	+F_{w}\right]_1^2=0\;, 
\end{equation}
where $\rho$, $u$, $p$ and $\gamma$ stand for density, velocity,
pressure and ratio of specific heats of the gas, $p_w$ and $F_w$ are
the magnetic pressure and energy flux and the brackets indicate the 
difference between quantities downstream and upstream of the subshock 
($[X]_1^2=X_2-X_1$). Some subtleties of the treatment of mass,
momentum and energy conservation in cosmic ray modified shocks are
extensively discussed in \cite{escape}. Here we only want to emphasize
that the contribution of the terms related to the pressure and energy
flux of the cosmic rays ($p_{c}$ and $F_c$) disappears when
considering the subshock (i.e. $p_c$ and $F_c$ are both continuous
across the subshock). 
Also the contribution to the pressure by the background magnetic field
vanishes, $\vec B_0$ being parallel to $\vec u$ and therefore
unaffected by any fluid compression.  The subshock Rankine-Hugoniot
relations are therefore not affected by the contribution of cosmic
rays, but they do take into account the presence of magnetic
turbulence, leading to a magnetized gaseous shock. 

In order to infer the magnetic field jump conditions, we use the
approach of \cite{sb71} and \cite{vs99} to describe the transmission 
and reflection coefficients appropriate for Alfv\'en waves at the 
shock surface. Following \cite{vs99}, we account for two upstream wave 
trains with helicities $H_c=\pm 1$, and for their respective
counterparts downstream. It is worth stressing that, in principle,
this part of the problem can be a complex one, because waves can be
transmitted and reflected also within the precursor, due to the
gradient in all quantities there. In general this could lead to 
isotropization of the turbulence and to the generation of a wave train 
propagating in the direction opposite to that directly excited by the
cosmic ray streaming (i.e. the one with $H_c=-1$).
However, we do not expect this effect to be very relevant because 
the shocks we are dealing with are not very modified, as we can check 
\emph{a posteriori}.
For these reasons we neglect turbulence isotropization in the 
precursor, but we include the transmission and reflection of Alfv\'en
waves at the subshock surface, which enter the jump conditions.

Let $\delta \vec B_{\mu}$ be a given mode, indicated by the subscript 
$\mu$, of the magnetic turbulence. 
We write the corresponding velocity perturbation simply as
\begin{equation}\label{disprel}
	\delta \vec u_\mu=-H_{c,\mu}\frac{\delta\vec B_\mu}{\sqrt{4\pi\rho}}\;.
\end{equation}
Neglecting the electric field contribution, which is of order
$u^2/c^2$, the magnetic pressure and the energy flux are 
respectively
\begin{equation}\label{pw}
	p_w=\frac{1}{8\pi}\left(\sum_\mu\delta \vec B_\mu\right)^2\;,
\end{equation}
\begin{equation}\label{Fw}
	F_w=\sum_\mu\frac{\delta \vec B_\mu^2}{4\pi}\left(u+H_{c,\mu} v_A\right)+
	\frac{\left(\sum_\mu\delta \vec B_\mu\right)^2}{8\pi}u\,,
\end{equation}
 where the sum over $\mu$ has to be intended as the sum over all
the wave modes present at a given position.

In obtaining these relations we explicitly used the fact that,
for Alfv\'en waves, $F_w$ in Eq.~\ref{Fw} has two contributions: the former is 
the normal component of the Poynting vector
$(\vec B\times \delta \vec u+\delta\vec B\times\vec u)\times\delta\vec B/4\pi$,  
while the latter represents the kinetic energy flux in transverse 
velocity, $\rho/2\delta\vec {u}^2 u$ (with $\delta u$ given by
Eq.~\ref{disprel}). 
If the turbulence manifests itself in forms other than resonant Alf\'en
waves, $p_w$ and $F_w$ may differ significantly.
We allow the generated waves to have both helicities, hence the sum 
over the modes should account, in the upstream region, for both
 forward- and backward-going waves, namely $\delta\vec B_{1\pm}$. 
Each of these modes, however, when advected behind the subshock has to
split into two waves with opposite helicities in order to satisfy the
Maxwell equations at the subshock \citep[see e.g][]{mck-w69,sb71}.
This fact leads to four different downstream modes, which can be
described by the introduction of proper transmission and reflection
coefficients. 
In this scenario a ``reflected'' (``transmitted'') wave is a wave with
a helicity opposite (equal) to the one of its upstream counterpart.
It is worth stressing that, the fluid being super-Alfv\'enic either
upstream or downstream, all the modes are actually advected with the
fluid, independent of their  direction of propagation.

These coefficients, $\mathcal T$ and $\mathcal R$ respectively, were
derived by \cite{mck-w69}, and for each upstream helicity $H_c=\pm 1$ read:
\begin{equation}\label{T}
	\mathcal
	T=\frac{\Rs+\sqrt{\Rs}}{2}\frac{M_{A1}+H_c}{M_{A1}+\sqrt{\Rs}H_c}  
	\simeq \frac{\Rs+\sqrt{\Rs}}{2}\;,
\end{equation}
\begin{equation}\label{R}
	\mathcal
	R=\frac{\Rs-\sqrt{\Rs}}{2}\frac{M_{A1}+H_c}{M_{A1}-\sqrt{\Rs}H_c} 
	\simeq \frac{\Rs-\sqrt{\Rs}}{2}\;. 
\end{equation}
Here $M_{A}= u/v_{A}$ is the Alfv\'enic Mach number, namely the ratio 
between fluid and Alfv\'en speed, which is of order 100 and more
for a typical supernova shock.
For each sign of $H_c$ we have $\delta B_2/\delta B_1=\mathcal
T+\mathcal R=\Rs$, and hence 
\begin{equation}\label{a2a1}
	p_{w2}=p_{w1} \Rs^2\;.
\end{equation}
Since the subshock can be viewed as a magnetized gas shock, as we
already stressed, the jump conditions found by \cite{vs99} for the
pressure and temperature hold also in our case and can be written respectively as:
\begin{equation}\label{p2p1}
\frac{p_2}{p_1}=\frac{(\gamma+1)\Rs-(\gamma-1)+(\gamma-1)(\Rs-1)\Delta}
	{\gamma+1-(\gamma-1)\Rs}\;,
\end{equation}
\begin{equation}\label{T2T1}
	\frac{T_2}{T_1}=\frac{p_2}{p_1}\frac{1}{\Rs}\,,
\end{equation}
with $\Delta$ defined as:
\begin{equation}	
\Delta=\frac{\Rs+1}{\Rs-1}\frac{\left[p_{w}\right]_1^2}{p_1}-\frac{2\Rs}
{\Rs-1}\frac{\left[F_w\right]_1^2}{p_1u_1}\;.
\end{equation}
Using the expressions in Eqs.~\ref{T} and \ref{R} for the transmitted
and reflected Alfv\'en waves, we find:
\begin{equation}\label{delta}
	\Delta=(\Rs-1)^2\frac{p_{w1}}{p_1}+\Rs~\frac{\delta\vec B_{1-}\cdot 
	\delta\vec B_{1+}}{2\pi p_1}\,.
\end{equation}

Following \cite{sb71} and \cite{vs99}, we assume that the two 
opposite-propagating waves carry magnetic fields $\delta\vec B_{1\pm}$ 
displaced in such a way that $\delta\vec B_{1-}\cdot\delta\vec B_{1+}=0$. 
This is not the most general possible configuration, but still it is
indeed expected to be the most common, since it describes situations 
in which there is only one wave train, or the two fields are 
reciprocally orthogonal, or even, on average, when the relative phase 
between the wave trains is arbitrary.

Hereafter we use quantities normalized to the values of ram-pressure
and velocity at upstream infinity: 
\begin{equation}
U(x)= \frac{u(x)}{u_0}\,\qquad 
\alpha(x)= \frac{(\sum_\mu\delta\vec B_\mu)^2}{8 \pi \rho_0u_0^2}\,\qquad
P(x)= \frac{p(x)}{\rho_0u_0^2}\,\qquad
\xi(x)= \frac{p_{c}(x)}{\rho_0u_0^2}\,.
\end{equation}
If the heating of the upstream plasma is due only to adiabatic 
compression, using the mass conservation $\rho(x)u(x)=\rho_0u_0$, the 
normalized plasma pressure can be written as
\begin{equation}\label{pgas}
	 P(x)=\frac{U(x)^{-\gamma}}{\gamma M_0^2}\;,
\end{equation}
where as usual $M_0$ is the sonic Mach number at upstream infinity.

Substituting Eq.~\ref{p2p1}, Eq.~\ref{delta} and the above expression
for $P(x)$ in the equation for momentum conservation, the compression
factors $\Rs$ and $\Rt$ are related through the equation 
\begin{equation}\label{rsrt}
\Rt^{\gamma+1}=\frac{M_0^2\Rs^\gamma}{2}\left[\frac{\gamma+1-\Rs(\gamma-1)}  
{1+\Lambda_B}\right],
\end{equation}
which is the same as the standard relation \citep[see e.g.][]{blasi2}
apart from the factor $(1+\Lambda_B)$, where 
\begin{equation}
\Lambda_B=W\left[1+\Rs\left(2/\gamma-1\right)\right]\;,
\end{equation}
and we have defined
\begin{equation}
	W=\alpha_1/P_1\,.
\end{equation}
It is clear that the net effect of the magnetic turbulence is to make
the fluid less compressible: if $W\simgt 1$, $\Rt$ may be considerably
reduced, while the pressure (and temperature) jump increases, according 
to Eq.~\ref{p2p1} (and \ref{T2T1}).
In \cite{jumpl} we showed, by means of purely hydrodynamical
considerations, and without any assumptions on the details of particle 
acceleration and magnetic field generation, that this is very likely
the case for the SNRs listed in Tab.~\ref{tab:SNR}.

As a final remark, we notice that if one naively assumed that downstream 
$F_{w2}=3u_2p_{w2}$, instead of using the appropriate transmission and 
reflection coefficients (which come from the need of satisfying
Maxwell equations at the subshock), one would obtain
\begin{equation}
	\Delta'=[(\Rs-1)^2-2\Rs] W < \Delta\,.
\end{equation}
Using $\Delta'$ rather than $\Delta$ leads to an incorrect estimate of
the pressure jump (Eq.~\ref{p2p1}), which in this case may even turn
out to be smaller than for an unmagnetized shock ($\Delta'<0$ for
$\Rs<3.73$). At the same time, one would have 
$\Lambda_B'=W\left[1+\Rs\left(3/\gamma-2\right)\right]$ in
Eq.~\ref{rsrt} and hence find a less marked decrease of $\Rt$.

\section{The kinetic self-consistent solution}\label{sec:model}

In this section we describe the calculations that lead to an exact solution for
the spectrum and the spatial distribution of the particles accelerated at a
non-linear astrophysical shock, including the generation of Alfv\'en waves by
the same particles and the dynamical reaction of both cosmic rays and magnetic
turbulence on the fluid. 
This method is based on the kinetic treatment of the problem in the 
stationary regime proposed by \cite{amato1,amato2}, which also allows 
for an arbitrary choice of spatial and momentum dependence of the
diffusion coefficient $D(x,p)$. 
In the following we consider Bohm diffusion in the self-generated 
magnetic field, i.e. we set
\begin{equation}
	D(x,p)=\frac{1}{3}c r_L(\delta B)=\frac{1}{3}c \frac{pc}{e B(x)}\,,
\end{equation}
where $r_L$ is the the Larmor radius of a particle of momentum $p$ in the local
amplified magnetic field $B(x)=\sqrt{8\pi \alpha(x)\rho_0u_0^2}$. Needless to
say that this form of the diffusion coefficient is basically only an ansatz and
whether Nature provides such a diffusion coefficient is at present not clear. 

The transport of accelerated particles consists of both advection and 
diffusion. The correct treatment of these processes in the limit of
small perturbation amplitudes ($\delta B\ll B_0$), also including the 
presence of both wave trains, are well known and can be found e.g. in 
the work by \cite{skillinga,skillingc}.
Unfortunately, in the case of strong magnetic field amplification a
full theory of cosmic ray transport is still missing, and even the 
definition of an effective wave velocity, $v_A$,  is troublesome.
A semi-analytical treatment, which is what we are interested in,
is only possible within the framework of quasi-linear theory. We adopt 
an Alfv\'en velocity defined as
\begin{equation}\label{valf}
	v_A(x)=\frac{B_0}{\sqrt{4\pi\rho(x)}},
\end{equation}
where $\rho(x)$ is the gas density in the precursor at the position $x$ and
$B_0$ is the strength of the unperturbed magnetic field. In the unlikely case
that the waves could keep their Alfv\'enic nature even in the strong turbulence
regime, this wave velocity would in fact be well defined.

As to the interactions between streaming particles and Alfv\'en waves,
we do not neglect the velocity of the scattering centers $v_A$ with
respect to the fluid velocity $u(x)$. This means that, in principle,
the compression ratios of the background fluid are different 
from the ones experienced by the scattering centers, which turn out to be
\begin{equation}\label{ssst}
	\Ss=\frac{u_1-v_{A1}}{u_2+v_{A2}}\qquad
 \St=\frac{u_0-v_{A0}}{u_2+v_{A2}}\simeq\frac{u_0}{u_2+v_{A2}}\,.
\end{equation}
In the recipes above we have implicitly assumed that upstream waves are produced
preferentially with $H_c=-1$.
 Behind the subshock, instead, the net velocity of the two
 opposite-propagating wave trains, when calculated in the downstream
 gas reference frame ($v_{A2}$), turns out to be in the same
 direction as the fluid one \cite[see also][]{bell78}.

In a typical SNR the condition $v_A\ll u$ usually holds, hence the
difference between ($\Ss,\St$) and ($\Rs,\Rt$) is expected not to be
very relevant. But if for some reason $M_A$ is small enough, the
compression ratios felt by the accelerated particles may be
significantly different with respect to the fluid ones, leading to a 
modified spectral slope, as already showed by \cite{bell78}.
This is why we retain $v_A$ in the calculations and check {\it a
posteriori} that in the cases considered this correction is not 
important. In Sec. \ref{sec:modif} we investigate the consequences of adopting
a different prescription for the velocity of the scattering centers.

From the kinetic point of view, cosmic rays are described by
their distribution function in phase space $f(\vec{x},\vec{p})$.
Keeping only the isotropic part (since $f(\vec{p})=f(p)+O(u^2/c^2)$)  and
recalling that the shock is non-relativistic, the diffusion-advection equation
for a one-dimensional shock reads:
\begin{equation}\label{conv-diff}
	\left[u(x)-v_A(x)\right]\frac{\de f(x,p)}{\de x}=
	\frac{\de}{\de x}\left[D(x,p)\frac{\de}{\de x}f(x,p)\right]
	+\frac{d\left[u(x)-v_A(x)\right]}{dx}\frac{p}{3}\frac{\de 
	f(x,p)}{\de p}+Q(x,p)\,,
\end{equation}
with $Q(x,p)$ the injection of particles in the accelerator. 
We assume that injection occurs only at the shock location ($x=0$)
and at momentum $\pj$, involving a fraction $\eta$ of the particles 
crossing the shock, such that
\begin{equation}
	Q(x,p)=\eta\frac{\rho_1u_1}{4\pi m_p \pj^2} \delta(p-\pj)\delta(x)\,.
\end{equation}

For the fraction $\eta$ of injected particles we adopt the recipe of 
\cite{bgv05}:
\begin{equation}\label{eta}
	\eta=\frac{4}{3\sqrt{\pi}}(\Ss-1)\psi^3 e^{-\psi^2}\,,
\end{equation}
which assumes that only particles with momentum $\pj\ge\psi p_{th,2}$ 
(i.e. $\psi$ times the thermal particles' momentum downstream) can be
accelerated. This recipe fits well within our self-consistent
approach because it only involves $p_{th,2}$ and $\Ss$, which are both 
outputs of the calculation, rather than free parameters. Moreover, it
self-limits the acceleration process, suppressing the injection when
particle acceleration is too efficient, in which case the large shock
modification leads to $\Ss\to 1$ and $\eta \to 0$. 
In the following, we consider values of $\psi$ between 3.5 and 4, 
corresponding to $\eta$ between $\sim 10^{-4}$ and $\sim 10^{-5}$. 

As shown by \cite{amato1,amato2} and then by \cite{bac07}, a very good 
approximation for the solution of Eq.~\ref{conv-diff}, $f(x,p)$, is
found in the form:
\begin{equation}
	f(x,p)=	f_1(p)\exp\left[\frac{\Ss-1}{\Ss}\, 
	  \frac{q(p) u_0}{3}
		\int_0^x dx'\frac{U(x')-V_A(x')}{D(x',p)}\right]\,,
\end{equation}
where $V_A=v_A/u_0$, $f_1=f(0,p)$ and $q(p)=-\frac{d\log 
f_1(p)}{d\log p}$ is the spectral slope at the shock location.
The above expression reduces to the correct distribution function in
the test particle limit and exactly satisfies the jump conditions at 
the subshock, as obtained by integrating Eq.~\ref{conv-diff} from 
$0^-$ to $0^+$.

As shown by \cite{blasi1}, $f_1(p)$ can be written as
\begin{equation}\label{f0p}
	f_1(p)=\left(\frac{3\St}{\St U_p(p)-1}\right)
	\frac{\eta\rho_1}{4\pi m_p \pj^3}
	\exp\left[-\int_{\pj}^p\frac{dp'}{p'}\frac{3\St U_p(p')}{\St 
	U_p(p')-1}\right]\,,
\end{equation}
where we defined 
\begin{equation}\label{Up}
	U_p(p)=U_1-V_{A1}-\frac{1}{f_1(p)}\int_{-\infty}^0 dx\ 
	f(x,p)\frac{d\left[U(x)-V_A(x)\right]}{dx}\,.
\end{equation}
The normalized pressure in cosmic rays is written in terms of this 
solution as
\begin{equation}\label{xicr}
	\xi(x)=\frac{4\pi}{3\rho_0u_0^2}\int_{\pj}^{\px}dp p^3 v(p) 
	f_1(p)\exp\left[
	\int_0^x dx'\frac{U(x')-V_A(x')}{x_p(x',p)}\right]\,,
\end{equation}
having defined
\begin{equation}	
	x_p(x,p)=\frac{3\Ss}{\Ss-1}\frac{D(x,p)}{u_0q(p)}, 
\end{equation}
and $\px$ as the maximum momentum achievable by the accelerated 
particles. We determine the latter self-consistently, using the
calculations by \cite{bac07} for cosmic ray modified shocks, 
assuming that the particles' maximum energy is limited by the acceleration time 
rather than by the size of the system. 
The time needed to accelerate particles up to $\px$ turns out to be 
\begin{equation}\label{Tpmax}
	\tau(\px)=\frac{3\Rt}{u_0^2}\int_{\pj}^{\px}\frac{dp'}{p'}
	\frac{1}{\Rt U_p(p')-1}
	\left[\Rt D_2(p')+\frac{u0}{f_1(p')}\int_{-\infty}^0 dx 
	f(x,p')\right]\,.
\end{equation}

Integrating by parts Eq. \ref{Up} it is possible to express $U_p(p)$ 
in terms of $U(x)$ and $x_p(x,p)$ alone:
\begin{equation}\label{Upp}
	U_p(p)=\int_{-\infty}^0 dx
	 \frac{\left[U(x)-V_A(x)\right]^2}{x_p(x,p)}\exp\left[
	\int_0^x dx'\frac{U(x')-V_A(x')}{x_p(x',p)}\right]\,.
\end{equation}
Then, differentiating Eq. \ref{xicr} with respect to $x$, we obtain
\begin{equation}\label{dxi}
	\frac{d\xi(x)}{dx}=\xi(x)\lambda(x) \left[U(x)-V_A(x)\right],
\end{equation}
where
\begin{equation}\label{lambda}
	\lambda(x)=\frac{\int_{\pj}^{\px} dp\,p^3\, v(p) 
	\frac{1}{x_p(x,p)}f(x,p)}
	{\int_{\pj}^{\px} dp\,p^3\, v(p) f(x,p)}\,.
\end{equation}
Finally, we need a relation which describes how the Alfv\'en waves are 
excited by the streaming cosmic rays and how the wave energy is 
transported in the precursor. Very generally, we can assume
$\alpha(x)$ to be a function of $\xi(x)$ and $U(x)$. Using the equation
of momentum conservation between a point $x$ in the precursor and
upstream infinity, it is therefore possible to write $\alpha(x)$ as a function
of $U(x)$ only (see e.g. Eq.~\ref{alfares} in the next section, where a discussion 
of magnetic field amplification due to resonant streaming instability will be presented.)

The nonlinear system defined by Eqs.~\ref{conv-diff}-\ref{lambda} can 
be solved, for a given age of the system, with three nested iterations. 
We guess a value for $\px$, as the starting point of the outermost 
cycle. Then, we fix a value for the ratio $\Rs/\Rt$, and derive the
corresponding $\Rs$ and $\Rt$ from Eq.~\ref{rsrt}. 
The equation for conservation of momentum,
\begin{equation}\label{eq:momentum}
U(x)+\xi(x)+\alpha(x)+P(x)=1+\frac{1}{\gamma M_0^2}\,,
\end{equation}
once it is evaluated in $0^-$, 
\begin{equation}\label{eq:xi1}
\frac{\Rs}{\Rt} + \frac{1}{\gamma M_0^2}\left(
 \frac{\Rt}{\Rs}\right)^\gamma  + \xi_1
 +\alpha_1=1+\frac{1}{\gamma M_0^2}\,,
\label{eq:prec}
\end{equation}
only involves $\xi_1$ and $\Rs/\Rt$, since $\alpha_1$ can be written as a function of $\Rs/\Rt$, 
as it will be shown in the next section (Eq.~\ref{alfares}). Eq.~\ref{eq:xi1} gives
the boundary condition $\xi_1$ for Eq.~\ref{dxi}. The latter is then
solved recursively: the solution at step $n$ is calculated using
$U(x)$ and $\lambda(x)$ at step $n-1$:
\begin{equation}
	\xi^{(n)}(x)=\xi_1\exp\left[\int_0^x
	 dx'\lambda^{(n-1)}(x')\left[U(x')-V_A(x')\right]^{(n-1)}\right]\,.
\end{equation}
The iteration on $n$ is carried out until convergence is reached. The
functions $U(x)$ (and thus $\Ss$ and $\St$), $U_p(p)$, $\lambda(x)$
and $f_1(p)$ are recalculated at every step, through Eq.~\ref{eq:momentum}, 
Eq.~\ref{Upp}, Eq.~\ref{lambda} and Eq.~\ref{f0p} respectively. 
In this way the value of $\xi(x=0)$ obtained by integration of 
$f_1(p)$ through Eq.~\ref{xicr} in general does not match the value of 
$\xi_1$ given by Eq.~\ref{eq:prec}, hence we restart the procedure
with a different ratio $\Rs/\Rt$ until this necessary condition is 
satisfied. 
Having thus found a solution, one is able to calculate the acceleration 
time which corresponds to it according to Eq.~\ref{Tpmax}.
Adjusting $\px$ obviously allows us to find the whole self-consistent 
solution for a fixed age of the SNR.

\section{Streaming instability}\label{sec:resSI}

Magnetic fields can be generated by streaming instability induced by cosmic
rays, although alternative models have also been proposed
\cite[see e.g.][]{gj07}. Streaming instability can be induced in a resonant
\citep{bell78} or non-resonant \citep{bell2004} way, depending on the type of 
interaction between particles and waves. In the former case, the unstable modes
are Alfv\'en waves, while in the latter case the modes are almost purely growing
modes and do not correspond to Alfv\'en waves. A satisfactory description of the
interaction of particles with these waves is missing, therefore it is very
problematic at the present time to describe cosmic ray diffusion in a
background of waves which are excited non-resonantly \citep[see][for some recent attempts]{ptu,reville}. 
Moreover, even the jump conditions at the subshock and in the precursor would be different from the ones typically adopted and it is not clear as yet which form the wave terms shuld
have. In the context of SNRs, \cite{amato3} showed that the non-resonant modes
are bound to be relevant only in the early stages of the evolution, while for
most of the history of the SNR, streaming instability should be dominated by the
excitation of resonant waves. For these reasons we chose to focus only on the
self-generation of resonant Alfv\'en waves, leaving other cases, including the
phenomenological approach of  \cite{bl2001}, for future work.

The stationary equation for the growth and transport of magnetic turbulence 
reads \cite[e.g.][]{mck-v82}:
\begin{equation}\label{transw0k}
	\frac{\partial \mathcal F_w(k,x)}{\partial x}=
	u(x)\frac{\partial \mathcal P_w(k,x)}{\partial x}+
	\sigma(k,x) \mathcal P_w(k,x)-\Gamma(k,x) \mathcal P_w(k,x)\;,
\end{equation}
where $\mathcal F_w$ and $\mathcal P_w$ are, respectively, the energy 
flux and pressure per unit logarithmic bandwidth of waves with
wavenumber $k$, $\sigma(k,x)$ is the rate at which the energy in magnetic
turbulence grows and $\Gamma(k,x)$ is the rate at which it is damped.
This equation is very general, but in the case of modified shocks one
should keep in mind the fact that Fourier analysis in $k$-space is
only accurate for wavenumbers such that $1/k$ remains appreciably smaller  
than the typical length scale of the precursor.

In the following we only consider resonant scattering between the 
accelerated particles and the magnetic turbulence, which gives for the
growth-rate of energy in Alfv\'en waves:
\begin{equation}\label{sigmak}
	\sigma(k,x)=\frac{4\pi}{3}\frac{v_A(x)}{\mathcal P_w(k,x)}
	\left[p^4v(p)\frac{\partial f(x,p)}{\partial x}\right]_{p=\bar{p}(k)}\,,
\end{equation}
where $\bar{p}(k)=eB/km_pc$ is the resonance condition 
\citep[see e.g.][for a derivation of this expression]{amato2}. 

Assuming no damping for the moment, and integrating Eq.~\ref{transw0k} in 
$k$-space we obtain:
\begin{equation}\label{transw0}
\frac{d F_w(x)}{d x}=u(x)\frac{d p_w(x)}{d x}+
	v_A \frac{d p_c(x)}{dx}\;.
\end{equation}
We further assume that only waves of one sign of helicity are
generated upstream and that $v_A\ll u$, so that	$F_w(x)\simeq
3u(x)p_w(x)$.
With these assumptions Eq.~\ref{transw0} becomes:
\begin{equation}\label{wtrans}
2U(x)\frac{d \alpha(x)}{dx}=V_A(x)\frac{d \xi(x)}{dx}
-3\alpha(x)\frac{d U(x)}{dx}\,.
\end{equation}
Provided the cosmic ray generation is efficient and that both $M_0$ and 
$M_A$ are much larger than 1, we can neglect the plasma and the magnetic 
pressure with respect to the kinetic and cosmic ray terms in 
Eq.~\ref{eq:momentum}, hence $\xi(x)\simeq 1-U(x)$
and Eq.~\ref{wtrans} can be solved analytically, returning
\begin{equation}\label{alfares}
	\alpha(x)=U(x)^{-3/2}\left[\alpha_0+\frac{1-U(x)^2}{4M_{A0}}\right]\,.
\end{equation}
We could use Eq.~\ref{wtrans} directly in our calculations, but it is
easy to check that the assumptions which lead to Eq.~\ref{alfares} are 
well satisfied in all cases of interest. The important physical
information to retain at this level is the enhancement of the
magnetic field due to adiabatic compression, which clearly shows in
Eq.~\ref{alfares} through the factor $U^{-3/2}$. 

In the following we consider the case $\alpha_0=0$, assuming that all 
the turbulence is generated via streaming instability, thus
\begin{equation}\label{resSI}
	\alpha(x)=\frac{1-U(x)^2}{4M_A(x)U(x)},
\end{equation}
where $M_A(x)=M_{A0}\sqrt{U(x)}$ is the local Alf\'enic Mach number.

For the sake of clarity, we stress that contributions of order $1/M_A$
are usually negligible, as in the calculation of the reflection and 
transmission coefficients or in the treatment of the magnetic energy
flux. The only exception to this rule is in the conservation equations 
for momentum and energy, where the magnetic terms, of order $1/M_A$,
are comparable to the ones pertaining the gas, usually of order $1/M_0^2$. 
Both these contributions are very relevant to the shock dynamics
because they affect the compressibility of the system. 
However, they are extremely small compared to the kinetic and cosmic ray 
energy, so that Eq.~\ref{alfares} is a very good approximation.

\section{Results}\label{sec:results}
In this section we show the results obtained through the algorithm
described in Sec.~\ref{sec:model}, and including only resonant
amplification of the magnetic field in the precursor, as described in
Sec.~\ref{sec:resSI}.

Here and in the following, unless specified otherwise, we assume a SNR age
of $1000\rmn{yr}$, a circumstellar density of $\rho_0=0.5m_p/\rmn{cm}^3$ and a background
magnetic field of $B_0=5\umu\rmn{G}$. The injection parameter $\psi$ is kept fixed:
$\psi=3.7$. We also consider two different circumstellar temperatures
$T_0=10^4$ and $10^6\degK$, which for $u_0=5900\rmn{km/s}$ correspond to
$M_0=500$ and $M_0=50$ respectively. This is done in order to investigate
different scenarios for shock propagation.

\begin{figure}
		\includegraphics[width=0.49\textwidth,height=180pt]{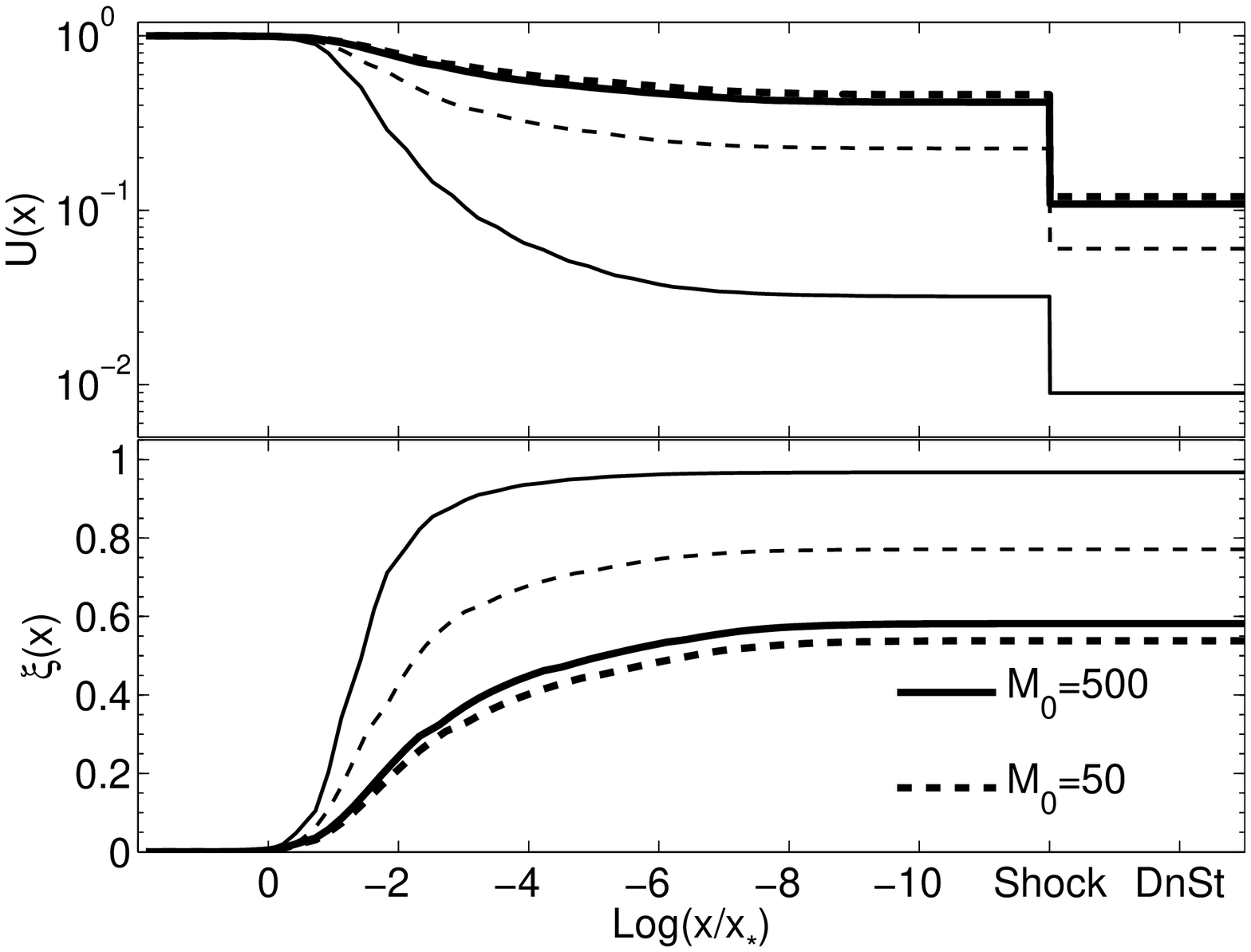}
		\includegraphics[width=0.49\textwidth,height=180pt]{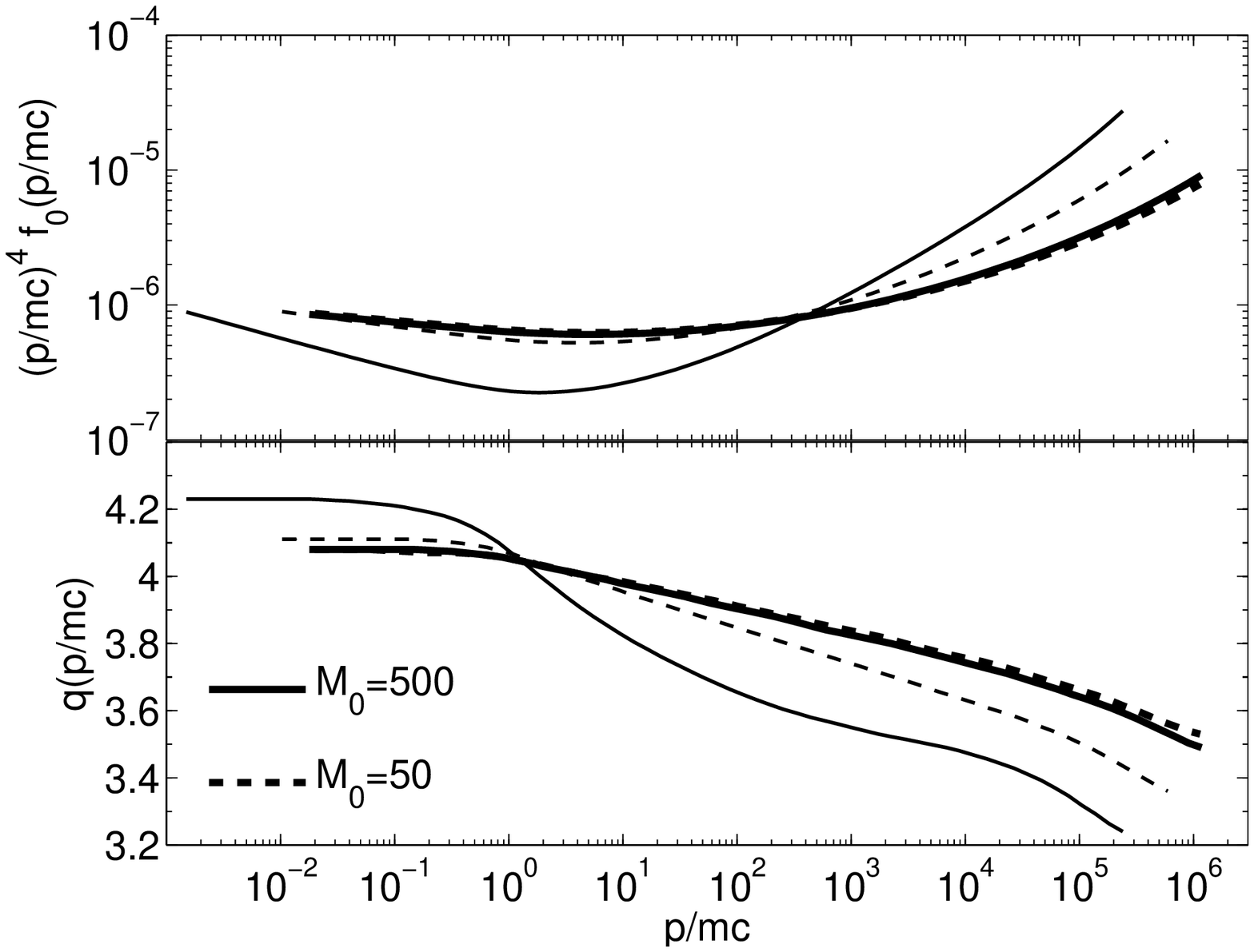}
	\caption{Left panel: velocity (\emph{top}) and cosmic ray pressure
	(\emph{bottom}) profiles.
	Right panel: momentum distribution (\emph{top}) and spectral slope 
	(\emph{bottom}) of the cosmic ray spectrum at the shock
	location. The \emph{thick lines} represent the solution found
	when the magnetization is correctly treated in the jump
	conditions, while the \emph{thin lines} show the solution
	found when it is ignored. Different line-types refer to
	different values of the Mach number $M_0$, as specified in the
	figure. The location upstream is in units of $x_*=-D(\px)/u_0$
	and is expressed in base-10 logarithm, while the downstream is taken as homogeneous.}
	\label{fig:Ucsispettro}
\end{figure}

In Fig.~\ref{fig:Ucsispettro} the velocity and cosmic ray pressure are
shown (\emph{top and bottom panels} on the left respectively) for
$M_0=50$ and $500$ (\emph{dashed and solid lines} respectively) with
or without inclusion of the magnetic feedback (\emph{thick or thin
 lines}). The values found for the relevant parameters in the same
cases are reported in Tab.~\ref{tab:err}.

The most striking effect that comes from the correct treatment of the
jump conditions for the magnetic field is indeed the reduced shock
modification, visible from the precursor profile: the total
compression ratio is found to be $\Rt\sim 9$ for both values of the
Mach number, while the prediction of standard non-linear theory would
be $\Rt\sim 112$ for $M_0=500$ and $\Rt\sim 17$ for $M_0=50$.

Clearly, the shallower precursor comes from the decrease of the
fraction of bulk energy that is converted into cosmic rays,
which however remains considerable: $\xi_1$ is reduced from more than 
90\% to around 50--60\%, for both values of $M_0$
(Fig.~\ref{fig:Ucsispettro}, left bottom panel).

The remaining fraction of bulk energy ends up being converted into heating
and energy of the turbulent magnetic field. We define the thermal emissivity as
\begin{equation}
	\varepsilon(x)\propto\rho(x)^2T(x)^{3/2}\propto U(x)^{-2}T(x)^{3/2}\,,
\end{equation}
and we report, in the last column of Tab.~\ref{tab:err}, the ratio
between its downstream value computed within modified shock theory
($\varepsilon_2$) and the prediction for the same shock in test
particle theory ($\varepsilon_{tp}$). The temperature (next to last
column in Tab.~\ref{tab:err}) and thermal emissivity downstream are
always much smaller than in the test-particle approximation, since 
a considerable fraction of the shock ram pressure is now going into 
particle acceleration and magnetic field amplification. 
Nevertheless, it is clear that the effect of 
the magnetized jump conditions is to enhance both quantities
(e.g. $T_2$ is enhanced by a factor $>100$ for $T_0=10^4\degK$), 
as a net result of the increased temperature jump $T_2/T_1$
(Eqs.~\ref{T2T1},\ref{p2p1}) and the reduced compressibility of 
the upstream plasma (lower $\Rt$).

We should recall that the compression ratios actually felt by 
cosmic rays depend on the relative velocity between them and 
the scattering centers, i.e. the Alfv\'en waves, according to
Eq.~\ref{ssst}. Nevertheless, when the Alfv\'en velocity is taken
according to Eq.~\ref{valf}, the discrepancy between $\Ss-\St$ and
$\Rs-\Rt$ is usually negligible, the difference in the plasma's and
cosmic rays' compression ratios being of order 1\% at most for the 
cases in Tab.~\ref{tab:err}.

As to the magnetic field, the values of $B_2$ in Tab.~\ref{tab:err} 
are in the correct range inferred from fits to X-ray observations of
SNRs, which indicate $B_2\approx400-500\umu\rmn{G}$ for SNRs with
$u_0$ as high as $5000-6000\rmn{km/s}$. This shows that amplification due
to streaming cosmic rays can actually account for such high magnetization
levels.

\begin{table*}
\begin{minipage}{150mm}
\caption{\label{tab:err} Comparison of the results obtained with and
 without inclusion of the magnetic feedback ($\Lambda_B$) for the
 shocks of Fig.~\ref{fig:Ucsispettro}.}
\begin{tabular}{ccccccccccc}
\hline
$T_0(\degK)$ & $\Lambda_B$ & $\xi_1$ & $\px(10^6\rmn{GeV/c})$ & $\Rs$ & $\Rt$ & 
$\Ss$ & $\St$ & $B_2(\umu\rmn{G})$ & $T_2(10^6 \degK)$ & $\varepsilon_2/\varepsilon_{tp}$\\
\hline
$10^4$ & No & 0.97 & 0.24& 3.58 & 112.1 & 3.43 & 108.7 & 645.8 & 0.88 & 0.030\\
$10^4$ & Yes & 0.58 & 1.17 & 3.84 & 9.22 & 3.79 & 9.12 & 463.9 &126.5 & 0.346\\
\hline
$10^6$ & No & 0.77 & 0.59 & 3.76 & 16.6 & 3.70 & 16.4 & 235.0 & 42.3 & 0.216\\
$10^6$ & Yes & 0.54 & 1.14 & 3.84 & 8.44 & 3.79 & 8.36 & 425.1 & 154.8 & 0.391\\
\hline
\end{tabular}
\end{minipage}
\end{table*}

Let us now consider the spectrum of the accelerated particles for the same
situations discussed above. These are plotted in the right top panel of 
Fig.~\ref{fig:Ucsispettro} (distribution function in momentum space, multiplied
by $p^4$), while the right lower panel shows the spectral slope $q(p)$ at the
shock location. For diffusion coefficients that increase with particle momentum,
higher energy particles stream further upstream of the shock than the
less energetic ones. In the case of a modified shock, this causes the 
former to experience compression factors even larger than 4, the
typical limit for strong shocks in the test particle regime. 
Since the spectral slope is basically determined by this effective compression
ratio, the resulting spectra are concave, softer at the lowest energies and
harder at the highest energies. It is intuitively clear that a smoothening of
the precursor (i.e. a reduction of the ratio $\Rt/\Rs$) results 
in a reduced concavity of the spectra as compared to the
standard prediction of nonlinear models for a given Mach number of the shock. 
This effect is evident from the comparison of thick and thin curves in
Fig.~\ref{fig:Ucsispettro}, where the dynamical reaction of the amplified field
is included: the particle spectrum is roughly close to $p^{-4}$ up to $10^3\rmn{GeV/c}$
and tends to be flatter above this energy, with a slowly changing slope.

Apart from this global concavity of the spectra, three main differences 
between the case with and without the correct jump conditions are worth 
being noticed: 

1) $p_{inj}$ increases due to the more efficient heating of the 
downstream plasma, and the fraction of particles injected into the 
acceleration mechanism (Eq.~\ref{eta}) also slightly increases (from 
$\eta=1.03\times 10^{-4}$ to $\eta=1.19 \times 10^{-4}$ for
$M_0=500$). This is consistent with the increase in $\Rs$ induced 
by the magnetic feedback.  

2) at the highest energies the spectra are somewhat steeper than
usually predicted for strongly modified shocks ($p^{-3.5}$ instead of
$p^{-3.2}$), but the total cosmic ray energy is still dominated by the highest
momenta; 

3) the maximum momentum achieved by non-thermal particles is increased 
by about a factor 2-5, and reaches a few times $10^6\rmn{GeV/c}$, as a result of the
smoothening of the precursor. We point out that during the free expansion phase
the shock velocity may be high enough to make the non resonant streaming instability substantially
contribute to the magnetic field amplification. This would reflect in a boost of
$\px$ for two concurring reasons: first, as a direct consequence of the
resulting decrease of the diffusion coefficient; second, as a consequence of an
enhanced magnetic feedback that would cause a further reduction of $\Rt$ (see
Eq.~\ref{Tpmax} for the time needed to accelerate a particle to $\px$).
For a typical remnant age of $\tau=1000\rmn{yr}$, the maximum energies we derive
appear to be in qualitative agreement with the knee in the proton spectrum as
observed e.g. by KASCADE \citep{antoni+05}.

\begin{table*}
\begin{minipage}{150mm}
\caption{\label{tab:res}Solution of DSA for different SNR
 environmental parameters.}
\begin{tabular}{cccccccccccc}
\hline
$T_0(\degK)$ & $n_0(\rmn{cm}^{-3})$ & $B_0(\umu\rmn{G})$ & $\xi_1$ &
$\px(10^6\rmn{GeV/c})$ & $\Rs$ & $\Rt$ & 
$W$ & $B_2(\umu\rmn{G})$ & $T_2 (10^6\degK)$\\
\hline
$10^4$ & 1 & 5 & 0.70 & 1.42 & 3.71 & 12.5 & 141.9 & 714.2 & 61.3\\
$10^4$ & 0.5 & 10 & 0.54 & 1.60 & 3.76 & 8.18 & 373.3 & 579.0 & 149.8\\
$10^4$ & 0.1 & 5 & 0.50 &  0.74 & 3.78 & 7.68 & 406.8 & 255.1 & 173.6\\
$10^4$ & 0.1 & 1 & 0.74 & 0.36 & 3.74 & 14.3 & 89.9 & 201.9 & 48.0\\
\hline
$10^6$ & 1 & 5 & 0.65 & 1.43 & 3.73 & 10.8 & 1.41 & 632.2 & 88.2\\
$10^6$ & 0.5 & 10 & 0.51 & 1.56 & 3.77 & 7.75 & 3.64 & 546.6 & 171.5\\
$10^6$ & 0.1 & 5 & 0.48 & 0.72 & 3.78 & 7.36 & 4.00 & 243.8 & 192.9\\
$10^6$ & 0.1 & 1 & 0.67 & 0.36 & 3.78 & 11.4 & 0.89 & 167.8 & 82.5\\
$10^6$ & 0.01& 5 & 0.12 & 0.15 & 3.94 & 4.50 & 4.41 & 54.7 & 586.0\\
$10^6$ & 0.01& 1 & 0.37 & 0.16 & 3.92 & 6.24 & 2.17 & 50.2 & 302.6\\
\hline
\end{tabular}
\end{minipage}
\end{table*}

Finally, we notice that our results show a weak dependence on the background
temperature $T_0$ (in the range $10^4-10^6\degK$), and therefore on the 
sonic Mach number, as long as this is much larger than 1. This
contrasts with the standard non linear prediction for the increase of
the shock modification as $\Rt\propto M_0^{3/4}$. The explanation
lies in the fact that the magnetic backreaction is much more 
effective for high $M_0$, since it is driven by 
\begin{equation}
	W=\frac{\alpha_1}{P_1}= \frac{\gamma}{4} \frac{M_0^2}{M_{A0}}
	\left(\frac{\Rs}{\Rt}\right)^{\gamma-3/2}
	\left[1-\left(\frac{\Rs}{\Rt}\right)^2\right]
	\propto\frac{u_0 B_0}{\sqrt{\rho_0} T_0}\,.
\end{equation}

In Tab.~\ref{tab:res} we show the results for several different choices of the
environmental parameters: it is clear that the magnetic feedback is never
negligible (namely we always have $W\gtrsim1$) and that the prediction for
the compression ratios is consistent with the values inferred from 
observations.
The case with $T_0=10^6\degK$ and low density $n_0=0.01\rmn{cm}^{-3}$, 
which may be representative of the hot phase  of the interstellar medium, 
does not lead to a shock as strongly  modified as in the other cases because 
in such an environment  an age older than $\sim 1000\rmn{yr}$ is needed 
to achieve a strong modification.

\subsection{The effect of the velocity of scattering centers on the spectrum}
\label{sec:modif}

As discussed in Sec.  \ref{sec:model} , the propagation of the accelerated
particles in the shock region is described by Eq. \ref{conv-diff}. The terms
$u(x)-v_A(x)$ describe the velocity of the scattering centers. Here $v_A$
represents the wave velocity, which, as discussed above, we assume to be the
Alfv\'en speed calculated in the background field $B_0$. Provided the waves
remain Alfv\'en waves even in the regime $\delta B/B\gg 1$ this result holds in
an exact way. It is however clear that this can only be considered as an ideal
case, in that turbulence may change such a picture in a considerable way. One
way in which the change may appear is in changing the wave speed. In this
section we investigate the effects on the spectrum of accelerated particles and
on the shock precursor which derive from calculating the Alfv\'en speed in the
local amplified field, namely:
\begin{equation}\label{valf1}
	v_A(x)=\frac{\delta B(x)}{\sqrt{4\pi\rho(x)}}.
\end{equation}
We stress that in our opinion this assumption is totally unjustified from the
physical point of view, and we use this case only as a toy model to illustrate
how sensitive the results can be to unknown non-linear effects. We notice
however that similar approaches have in fact been adopted, for instance by
\cite{poster} and also, in some form, by \cite{bl2001}. 

The net effect of this apparently harmless assumption is that the velocity of
the scattering centers is greatly enhanced and this affects in an substantial
way the effective compression factor at the subshock and therefore the
spectrum, as was already pointed out by \cite{bell78} in the context of test
particle theory.

\begin{figure}
		\includegraphics[width=0.49\textwidth,height=180pt]{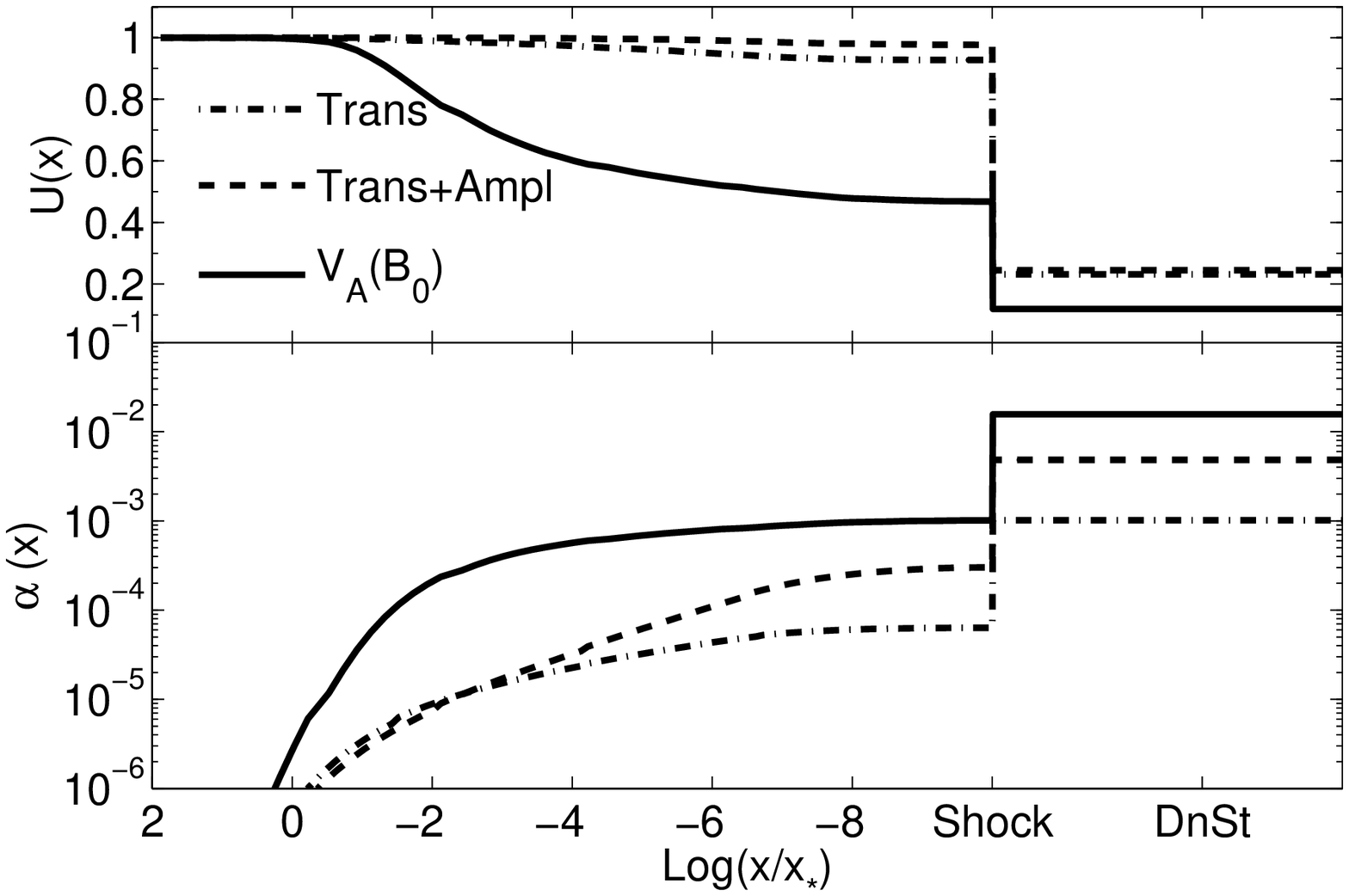}
		\includegraphics[width=0.49\textwidth,height=180pt]{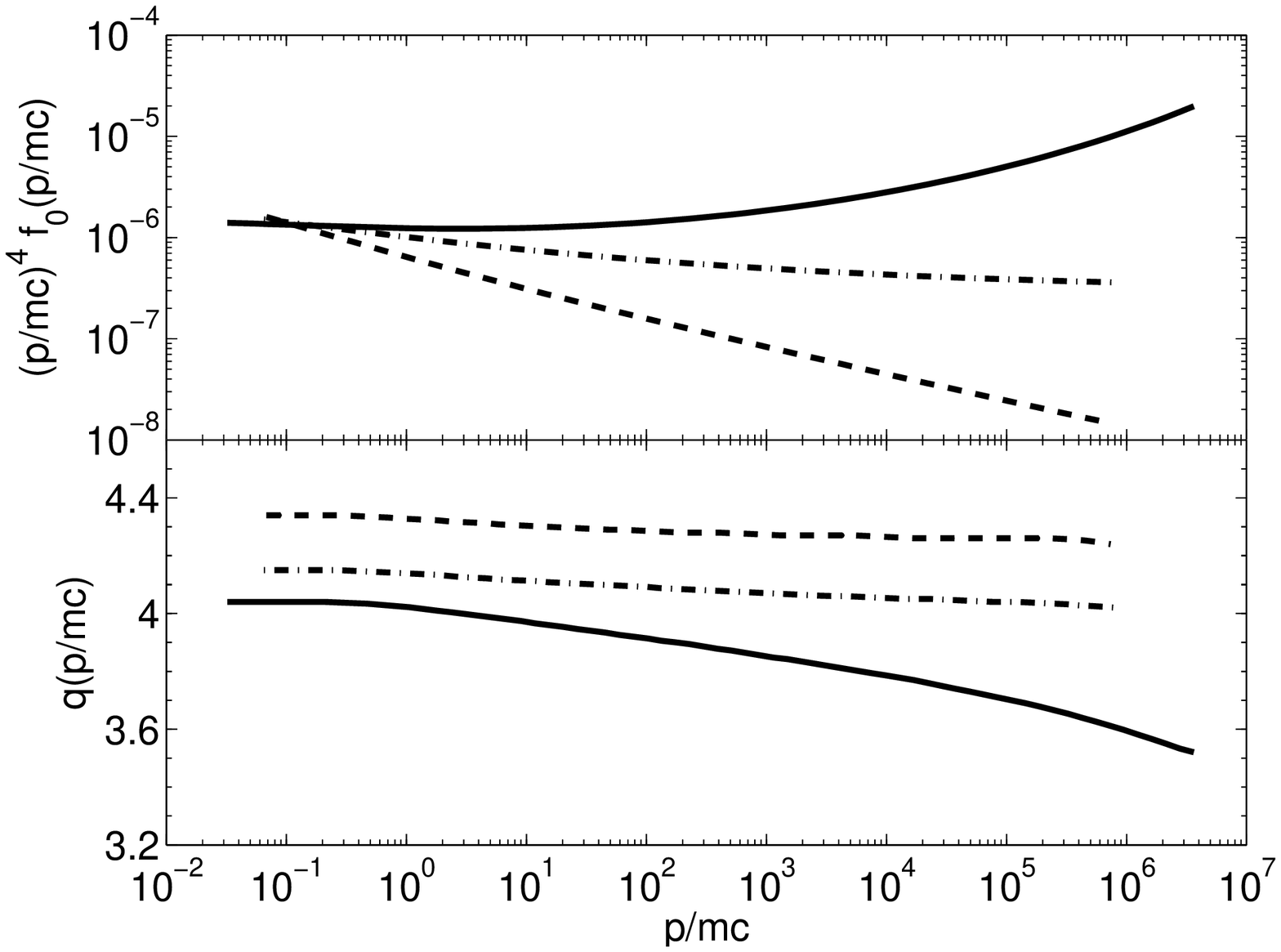}
	\caption{Left panel: velocity (\emph{top}) and normalized magnetic
	energy density in the shock region (\emph{bottom}) .
	Right panel: momentum distribution (\emph{top}) and spectral slope 
	(\emph{botttom}) of the cosmic ray spectrum at the shock
	location. The \emph{solid lines} refer to the solution found
	when the Alfv\'en speed is calculated in the background
	field. The \emph{dash-dotted lines} refer to the case in which the
	modified Alfv\'en speed is used only in the transport equation, while
	the \emph{dashed lines} refer to the case in which the modified Alfv\'en
	speed is used both in the transport equation and in the growth rate of
	unstable modes.}

	\label{fig:vAmod}
\end{figure}

In order to illustrate this effect in a quantitative way we run our calculations
for $p_{inj}=3.7 p_{th,2}$, $M_0=250$,  $T_0=10^5$ K, $n_0=0.5\rmn{cm}^{-3}$,
$B_0=5\umu\rmn{G}$ and an age of the system of $\sim1000\rmn{yr}$. The results are
illustrated in Fig. \ref{fig:vAmod}. The solid lines are obtained using the
Alfv\'en speed calculated in the background magnetic field $B_0$, as in the
previous section. In this case the precursor is very evident (top left panel in
Fig. \ref{fig:vAmod}) and the spectrum of accelerated particles is concave (top
right panel in Fig. \ref{fig:vAmod}).

The dash-dotted lines refer to the case in which the Alfv\'en speed is
calculated using the amplified magnetic field, but this definition is used only
in the transport equation, while the rate of growth of unstable waves is left
unchanged. As a consequence of the reduced effective compression ratio of the
scattering centers' velocity felt by accelerated particles, the spectrum
becomes softer, as illustrated in the right panels of Fig. \ref{fig:vAmod}. In
fact one can see that the spectrum becomes even steeper than $p^{-4}$ at all
momenta. When the modified Alfv\'en speed is used also in the growth rate
(dashed lines), the effect is even more dramatic and the spectrum has a slope
$\sim 4.3$ at all momenta and the concavity is hardly visible. Moreover, 
the amplified field becomes smaller in the latter two cases, which also results
in lower maximum energy of the accelerated particles, as visible in the right
panels of Fig. \ref{fig:vAmod}.

Despite the fact that the assumption of very large $v_A$ could be unphysical,
and certainly not well justified from the theoretical point of view, we cannot
refrain from being very concerned by the dependence of the results
on the value of the effective velocity of the scattering centers. 
On the other hand, having larger fields does not necessarily imply faster
scattering centers. For instance, the non-resonant waves discussed by 
\cite{bell2004}
and \cite{amato3} are almost purely growing modes and one could expect that they
may be almost stationary in the fluid frame. In addition, one should keep in
mind that if the velocity of the waves becomes too high, they may generate
shocks in the background plasma and damp their energy on it, so that those
waves do not contribute to the scattering of particles. 

\section{The (reduced) role of turbulent heating}\label{sec:TH}

Here we address another very important issue, already introduced in the work by
\cite{jumpl}. The correct inclusion of the turbulent magnetic field in the
calculation of the jump conditions very naturally leads to values of $\Rt$ that
allow to fit the X-ray observations in a totally new way in the perspective of
cosmic ray modified shocks, i.e. without invoking the presence of additional gas
heating mechanisms. 

It is well known that the shock dynamics, and in turn the particle acceleration
process, is very sensitive to the presence of non-adiabatic gas heating, since a
hotter upstream plasma is naturally less compressible. The low compression
ratios inferred from observations have been often explained by invoking
mechanisms of non-adiabatic heating in the precursor, such as \emph{Alfv\'en
heating} \citep{v-mck81,mck-v82} and \emph{acoustic instability}
\citep{df86,wagner+07}.

Acoustic instability develops when sound waves propagate in the pressure
gradient induced by cosmic rays in the shock precursor. The instability results
in the formation of weak shocks upstream, which cause heating in the precursor,
thereby reducing the compressibility of the plasma. \cite{wagner+07} showed
that there is however a range of steady state solutions characterized
by moderate cosmic-ray acceleration and compression ratios significantly larger 
than 7.

{\it Alfv\'en heating} (also called turbulent heating) is a generic expression
which is supposed, at least in principle, to apply to any damping mechanism for
Alfv\'en waves, and may
result for instance from ion-neutral damping or non-linear Landau damping,
depending on the ionization level of the background plasma. Although the
initial mathematical approach of \cite{mck-v82} was based on the assumption of 
non-linear Landau damping, the formalism was generic enough that it could be
adapted to any damping mechanism, and this is indeed what happened.
The common formulation of the Alfv\'en heating assumes that some fraction of
the energy in the form of waves is damped into the thermal energy of the
background plasma, independently of the details. Notice that this formalism
does not distinguish among modes with different wavenumber $k$, therefore this
type of calculations is intrinsically insensitive to the spectrum of turbulence
and should not be used to infer information on the shape of the diffusion
coefficient. Notice also that whenever applied to the case of non-linear Landau
damping, the mechanism is effective only when $u_0\ll 4000\rmn{km/s}
(T_0/5\times 10^5 K)^{1/2}$\citep{v-mck81}. It is not obvious that this
condition holds in the young SNRs listed above, since typically $u_0\sim
3000-6000\rmn{km/s}$.

A few other points are worth being mentioned: 1) Alfv\'en heating was first
introduced in order to avoid that the amplified magnetic field could reach the
nonlinear regime. On the other hand the same formalism is now used even in
situations in which $\delta B\gg B$. Much care should be taken in using these
expressions in the nonlinear regime. 2) The formalism introduced by
\cite{v-mck81} and  \cite{mck-v82}, and later adopted by virtually all authors
willing to include Alfv\'en heating, is based on the implicit assumption that
there is a rapid damping of all energy of waves onto the background plasma (as
stressed above, this was done in order to avoid excessive magnetic field
amplication). This inhibits the generation of magnetic field, so that large
magnetic fluctuations become unfeasible. In other words, the standard treatment
of Alfv\'en heating, which corresponds to  having $\sigma(x)=\Gamma(x)$ in
Eq.~\ref{transw0k}, is incompatible with having large values of the turbulent
field at the shock, as we show below in a quantitative way.

In the following we assume that the damping rate is limited to a fraction of the
growth rate, namely that
\begin{equation}\label{zeta}
	\Gamma(x)=\zeta \sigma(x)\,.
\end{equation}

\begin{table*}
\begin{minipage}{150mm}
\caption{\label{tab:TH}Alfv\'en heating effects}
\begin{tabular}{ccccccccccc}
\hline
$\zeta$ & $\xi_1$ & $\px(10^6\rmn{GeV/c})$ & $\Rs$ & $\Rt$ & $B_1/B_0$ & W &
$B_2(\umu\rmn{G})$ & $T_2 (10^6\degK)$ & $\varepsilon_2/\varepsilon_{tp}$\\
\hline
0 	& 0.60 & 1.17 & 3.76 & 9.52 & 25.3 & 1.941 & 475.6 & 114.6 & 0.317 \\
0.5 & 0.66 & 0.84 & 3.65 & 10.96& 20.8 & 0.390 & 379.6 & 132.6 & 0.523 \\
0.8 & 0.65 & 0.53 & 3.68 & 10.76& 12.8 & 0.115 & 232.5 & 128.3 & 0.480 \\
0.99& 0.55 & 0.12 & 3.85 & 8.69 &  2.26& 0.005 & 43.5 & 162.2 & 0.553 \\
\hline
\end{tabular}

\medskip
Solutions found for a shock with $T_0=10^6\degK$, $M_0=50$, 
$\rho_0=1m_p/\rmn{cm}^3$ and $B_0=5\umu\rmn{G}$ when the turbulent heating is
taken into account according to Eq.~\ref{rsrtTH}, for various
$\zeta<1$ (see Eq.~\ref{zeta}). The last column shows the downstream
thermal emissivity $\varepsilon_2$, normalized to the value it would
have for the same shock in the test particle regime. 
\end{minipage}
\end{table*}

An equation describing the Alfv\'en heating of the precursor can be
obtained by taking the derivative of the equation for the conservation
of energy together with the equations of transport of waves and cosmic
rays \citep[see the derivation of Eq.~9 in][]{mck-v82}. Under the above
assumptions we obtain: 
\begin{equation}\label{consen}
	\frac{\partial}{\partial x}\left[P_{TH}(x)U(x)^\gamma\right]=
	\zeta(\gamma-1) V_A(x)U(x)^{\gamma-1}
	\frac{\partial\xi(x)}{\partial x}\,,
\end{equation}
and $P_{TH}$ is now the pressure of the plasma including the effects of the turbulent heating.
Clearly, in the limit of adiabatic evolution of the precursor, one has 
$\zeta=0$ and Eq.~\ref{pgas} is recovered. 

It has been shown \citep{be99, amato2} that a good approximation to
the solution of Eq.~\ref{consen} is
\begin{equation}\label{pgasTH}
	P_{TH}(x)\simeq P(x)\left\{1+\zeta(\gamma-1)\frac{M_0^2}{M_A}
	\left[1-U(x)^\gamma\right]\right\}\,.
\end{equation}
where $P(x)$ is the plasma pressure as calculated taking into
account only adiabatic compression in the precursor, Eq.~\ref{pgas}.
This expression, which serves as an equation of state for the gas in the
presence of effective Alfv\'en heating, reduces to the standard Eq.~50 
of \cite{be99} for $\zeta=1$, while, for $\zeta<1$ the damping of the
waves is mitigated and an effective amplification of the magnetic
field is allowed.

The change in the equation of state of the gas also manifests itself
in the shock dynamics. The new relation between the compression ratios 
$\Rs$ and $\Rt$ reads
\begin{equation}\label{rsrtTH}
\Rt^{\gamma+1}=\frac{M_0^2\Rs^\gamma}{2}\left[\frac{\gamma+1-\Rs(\gamma-1)}  
{(1+\Lambda_B)(1+\Lambda_{TH})}\right]\,,
\end{equation}
where we have introduced
\begin{equation}
	\Lambda_{TH}=\zeta(\gamma-1)\frac{M_0^2}{M_A}
	\left[1-\left(\frac{\Rs}{\Rt}\right)^\gamma\right]\, ,
\end{equation}
with a notation which allows a direct comparison between the effects
of magnetic feedback and Alfv\'en heating.

It is widely known that the inclusion of Alfv\'en heating has an
important impact on the total compression ratio, changing its scaling 
with the Mach number from $M_0^{3/4}$ to $M_A^{3/8}$ \citep[see e.g.][]{be99}. 
However the situation is very different when the correct magnetized jump 
conditions are taken into account.
In Tab.~\ref{tab:TH}, we report, for different values of
$\zeta=0,0.5,0.8,0.99$, the solutions of the problem including the
full treatment of growth and damping of Alfv\'en waves according to the approximate 
analytical solution of Eq.~\ref{transw0k} along with the prescription of Eq.~\ref{zeta}, namely
\begin{equation}
	\alpha_{TH}(x)=(1-\zeta)~U(x)^{-3/2}\left[\alpha_0+\frac{1-U(x)^2}{4M_{A0}}\right]\,.
\end{equation}

It is clear that the increasing relevance of
Alfv\'en heating ($\zeta$ approaching 1) does not lead, in this
approach, to a smoother precursor (larger value of $\Rs/\Rt$). In
fact, the energy transfer from the waves to the plasma, while heating
the plasma and reducing its compressibility, is also accompanied by a 
decrease of $W$, the ratio of magnetic/plasma pressure. The latter is the
parameter which controls the magnetic feedback, which is then reduced.
The net effect is a slight increase of $\Rt$ for intermediate values
of $\zeta=0.5-0.8$. Only if $\zeta$ is very close to 1 the shock is
less modified than the adiabatic solution with magnetic backreaction, 
but even this effect is rather limited, since the decrease of $\Rt$ is
only about $\sim 10\%$  for the case $\zeta=0.99$.

The main effect of the inclusion of the Alfv\'en heating is instead a
significant reduction of the magnetic field (more than a factor 10 between 
$\zeta=0$ and $\zeta=0.99$), which also leads to a correspondingly
lower $\px$ ($\px\simeq 10^5\rmn{GeV/c}$ for $\zeta=0.99$).

We notice that the downstream temperature is affected in two different 
ways by the correct jump conditions and by the turbulent heating: the 
former provides an increase in the jump $T_2/T_1$ proportional to $W$, 
while the latter results in a higher $T_1$. The interplay between the
two effects produces the non-monotonic trend of $T_2$ and of the
thermal emissivity $\varepsilon_2$ (shown in the last columns of 
Tab.~\ref{tab:TH}). The latter is slightly increased by the presence
of non-adiabatic heating, with a value around 0.5$\varepsilon_{tp}$, 
basically for all choices of $\zeta>0$.

A recent investigation on the role of turbulent heating on the properties
of cosmic ray modified shocks has been carried out by \cite{vladimirov08}. 
These authors adopt a different recipe for particle injection into the 
acceleration process (which is a poorly understood issue in any case), 
that leads to a dependence of the fraction of injected particles on the 
temperature immediately upstream of the shock. As a consequence they find 
that dissipation of magnetic turbulence into heat upstream of the shock 
boosts the injection by a large factor and modifies the cosmic ray spectrum
at low energies, although it does not significantly affect the overall 
acceleration efficiency and the high energy part of the spectrum unless the
the fraction of turbulence energy that is transferred to heat gets close 
to 100\%. If the latter is the case, on the contrary, also 
\cite{vladimirov08} observe a considerable decrease of both the magnetic field
strength and $p_{max}$, in agreement with our results.   

Before concluding this section, we summarize our main conclusions concerning the
effects of turbulent heating:
\begin{itemize}
\item  if turbulent heating is in the form of nonlinear Landau damping,
it is suppressed for $u_0\gg 4000\rmn{km/s} (T_0/5\times 10^5\degK)^{1/2}$
\citep{v-mck81}, thus it could be important only if the circumstellar
temperature is rather high (for instance because the shock is expanding in the
pre-supernova stellar wind, or because of a dominance of the hot coronal
gas phase) and the shock has already slowed down significantly; 
\item if the rate of damping of the waves is too close to that of growth
($\zeta\sim1$), the magnetic field amplification is heavily suppressed 
and explaining the large levels of magnetization inferred from X-ray
observations becomes very challenging;
\item if the damping is effective but still allowing sufficient magnetic field
amplification, the smoothening of the precursor is roughly unchanged with
respect to the results obtained in the case of magnetic feedback alone. 
\end{itemize}

\section{Conclusions}\label{sec:conclusions}

The possibility that the narrow filaments detected in X-rays may result from
severe synchrotron losses of high energy electrons has sparked much attention
on the issue of magnetic field amplification at shock fronts, since the
inferred fields are $\sim 100$ times larger than the typical interstellar
ones. The importance of such magnetic fields for the origin of cosmic rays
is immediate: if the turbulent field leads to Bohm diffusion, it is easy to
show that particle acceleration at shock fronts may lead to the generation of
protons with energy close to the knee. 

In a previous paper \citep{jumpl} we investigated the hydrodynamical
consequences of having a large magnetic field amplified by cosmic ray
streaming upstream of the shock and we demonstrated that when the magnetic
pressure equals or overcomes the pressure of the background plasma, a
condition easy to realize, the compressibility of the plasma is reduced, 
so to decrease the compression factor of the cosmic ray modified shock and 
smoothen its precursor. 

In the present paper we went beyond the hydrodynamical picture and we solved
the combined system of conservation equations, cosmic ray diffusion-convection 
equation, and equation for the magnetic field amplification in order to 
investigate the effects of the precursor smoothening on the spectrum 
of accelerated particles. We find that resonant streaming instability 
is sufficient to amplify a parallel pre-existing magnetic field to levels 
which compare well with X-ray observations, if the latter are interpreted 
as a result of strong synchrotron losses. 

In these circumstances, we also confirm the crucial role of the dynamical 
feedback of the magnetic field, which leads to total compression factors 
around $\sim 7-10$ (to be compared with the typical predictions of 
standard NLDSA, which gives $R_{tot}\sim 20-100$), in good agreement 
with the values suggested by observations and by fits to the multifrequency 
spectrum of several SNRs \citep{V+05,warren}.
The reduced compression in the precursor does not inhibit particle
acceleration, in fact a fraction 50-60\% of the ram pressure at the shock is
converted into accelerated particles, for parameters typical of a SNR in the
Sedov phase. 

The kinetic calculation of particle acceleration in the nonlinear regime was
carried out using the approach of \cite{amato1,amato2}. The effects of resonant 
streaming instability were treated as in \cite{skillingc,bell78}, but
including the presence of the precursor which implies magnetic field
compression in addition to amplification. We neglected here the possibility of
having non-resonant amplification \citep{bell2004}, for several reasons: 1)
the relation between pressure and energy flux of these waves is not yet well
defined; 2) the diffusion properties of charged particles in turbulence
at wavelengths much shorter than the Larmor radius of particles are not known
as yet, and need a dedicated effort of investigation; 3) as showed by
\cite{amato3}, the non resonant modes are likely to be especially important in
the free expansion phase of a SNR and in the early stages of the Sedov phase.

Perhaps not surprisingly, we find that the effect of magnetic feedback on the
spectrum of accelerated particles is that of reducing the concavity. This
result can be easily explained based on the reduced compressibility in the
shock precursor, caused by the magnetic pressure. The slope of the spectrum
typically remains close to $4$ at energies smaller than $\sim 1$ TeV, while
being flatter at higher energies (Fig. \ref{fig:Ucsispettro}). We also discuss
the effect of the amplified field on the velocity of the scattering centers,
which causes a further steepening of the spectrum, leading it to get closer
to a power law. We find that the spectrum of accelerated particles becomes even
steeper than $p^{-4}$ if the velocity of the scattering centers is assumed to
be the Alfv\'en speed in the amplified field. Although this maybe an unphysical
assumption, it serves the scope of showing that the results of calculations
carried out in this strongly nonlinear regime may well be affected by details
which at the present time we are unable to control. 

The smoother precursor induced by the magnetic feedback also implies a larger
value of the maximum achievable momentum for the accelerated particles. We
recall that, as showed by \cite{bac07}, a strong precursor leads to lower the
maximum momentum since particles feel a smaller mean fluid speed and therefore
a longer acceleration time. Using the acceleration time for modified shocks as
calculated by \cite{bac07}, we find that for a $1000\rmn{yr}$ old remnant the
maximum energy can be as high as $10^6\rmn{GeV}$, a factor 5-10 larger than one
could find without including the magnetic backreaction. In principle even
larger maximum energies can be achieved by taking into account the non-resonant
streaming instability in the early phases of the SNR evolution.

 Our entire analysis was carried out for the case of a shock
  propagating in the direction parallel to the ambient magnetic
  field. The amplification of the background field due to cosmic ray
  streaming soon changes this situation, in that the magnetic field
  upstream becomes highly turbulent. Changing the field obliquity
  should not change much the picture discussed in the present paper as
  long as the projection of the shock velocity along the field lines
  remains supersonic. If the supernova shock propagates in a medium in
  which the background field is mainly oriented in a given direction,
  there will be locations at which the shock is quasi-parallel and
  others at which it is quasi-perpendicular. The implications for the
  morphology of the non-thermal emission require a dedicated
  investigation, in that the quasi-perpendicular regions could be
  expected to be bright because of the more efficient perpendicular
  configuration, but quasi-parallel regions could be expected to be
  bright because of more efficient magnetic field amplification. Clearly
  the degree to which these expectations may be true depends on many
  details which at present are not under control, although
  observations of specific astrophysical objects (e.g. SN 1006) will
  provide us with important information on this issue.

Finally, we investigated the effect of turbulent heating on the shock
modification. We generalized the formalism introduced by \cite{v-mck81} and 
\cite{mck-v82} in order to account for the possibility that a finite fraction
of the wave energy is damped on the background plasma. We find that, although
in general the heating of the upstream plasma results in a decrease of the
compressibility, the effect is at most competitive, if not subdominant, compared
with the dynamical feedback of the amplified magnetic field on the plasma. 
When turbulent heating is indeed dominant, a fraction very close to unity of
the wave energy is transformed in thermal energy, and the main effect is that of
suppressing the growth of the magnetic field, so that it becomes challenging to
explain X-ray observations in terms of severe synchrotron losses. Moreover in
this case the maximum energy of accelerated particles is drastically reduced
and it becomes much smaller than the knee. 

It is worth stressing that while the magnetic feedback on the background plasma
is based on well known physics, the turbulent heating is treated at best in a
very phenomenological way, with little attention to the specific physical
processes that may be responsible of the non-adiabatic heating and on how modes
with different wavelengths are damped. In these circumstances we think that the
effect of magnetic feedback is much more solidly assessed and should be
considered as the chief mechanism for reducing the compression in cosmic ray
modified shock waves with evidences for amplified magnetic fields. 

On the other hand, the role of turbulent heating may become much more important
for older remnants, when the shock velocity drops below $\sim 2000
\rmn{km/s}$. At these times, the amplification of the magnetic field may have
become less important from the dynamical point of view, so that the magnetic
feedback is also less important. These stages, as discussed by \cite{pz05}
play an important role in determining the spectrum of cosmic rays observed at
the Earth, at least at energies much lower than the knee energy.  

\section*{Acknowledgments}
The authors are grateful to the anonymous referee for a careful reading of the paper and for providing detailed comments that helped improving the manuscript. This work  was partially supported by PRIN-2006, by ASI through contract ASI-INAF I/088/06/0 and (for PB) by the US DOE and by NASA grant NAG5-10842. Fermilab is operated by Fermi Research Alliance, LLC under Contract No.DE-AC02-07CH11359 with the United States DOE.

\label{lastpage}

\end{document}